\documentstyle[aasms4,11pt]{article} 

\begin{document}
\title{DYNAMICAL MASS OF TYPE 2 SEYFERT  NUCLEI}
\author{\sc Shingo Nishiura, and Yoshiaki Taniguchi}
\vspace {1cm}
\affil{Astronomical Institute, Tohoku University, Aoba, Sendai, 980-77, Japan;
nishiura@astroa.astr.tohoku.ac.jp, tani@astroa.astr.tohoku.ac.jp}

\begin{abstract}
We have  derived the masses of central objects ($M_{\rm BH}$) of nine
type 2 Seyfert nuclei using the observational
properties of the {\it hidden} broad H$\beta$ emission.
We obtain the average dynamical mass, 
log$(M_{\rm BH} / M_\odot) \simeq 8.00 \pm 0.51 - 0.475 {\rm log}(\tau_{\rm es}/1)$
where  $\tau_{\rm es}$ is the optical depth for electron scattering.
If  $\tau_{\rm es} \sim 1$, this  average mass is 
almost comparable with those of type 1 Seyfert nuclei.
However, if  $\tau_{\rm es} \ll 1$, as is usually considered,
the average mass of type 2 Seyfert nuclei may be more massive than that of type 1s.
We discuss implications for issues concerning both the current 
unified model of Seyfert nuclei and physical conditions of the electron
scattering regions.
\end{abstract}


\section{INTRODUCTION} 

It is generally considered that active galactic nuclei (AGNs) are powered by
single, accreting supermassive black holes (e.g., Rees 1984; Blandford 1990).
According to this scenario, the accretion rate onto a black hole is an
important parameter
to explain the huge luminosity of AGNs.  However, since the luminosity released
from this central engine is proportional to the mass of the black hole
[ie., the Eddington luminosity, $L_{\rm Edd} \sim 10^{46} (M_{\rm BH}/10^8
M_\odot)$ erg s$^{-1}$ where $M_{\rm BH}$ is the black hole mass], 
the mass itself is considered as another important  parameter (Blandford 1990).
Relationships between mass and luminosity of AGNs provide important
information about the nature of central engines (e.g., Wandel \& Yahil 1985;
Padovani \& Rafanelli 1988; Padovani 1989;
Koratkar \& Gaskell 1991b). Further, the mass
function of nuclei may place  constraints on 
the formation and evolution of supermassive black holes in the universe
(Padovani, Burg, \& Edelson 1990; Haehnelt \& Rees 1993).
Therefore, the mass of AGNs is of fundamental importance in understanding
the AGN phenomena.

In order to estimate the nuclear mass,
the so-called dynamical method has been often used
(Dibai 1981, 1984; Wandel $\&$ Yahil 1985; Wandel \&
Mushotzky 1986; Joly et al. 1985; Reshetnikov 1987; Padovani \& Rafanelli
1988; Padovani, Burg, \& Edelson 1990;  Koratkar \& Gaskell 1991b).
If the gas motion in a broad emission-line region (BLR) is dominated
by the gravitational force exerted by the central massive object, the
line width can be used to estimate the mass of the central object
given the radial distance of the BLR (Woltier 1959; Setti \& Woltier 1966).
Since the recent elaborate monitoring observations of AGNs have shown
that the gas motion in the BLRs is almost
dominated by the gravitation (Gaskell 1988; Koratkar \& Gaskell 1991a, 1991c;
Clavel et al. 1991; Peterson 1993; Robinson 1994; Korista et al. 1995;
Wanders et al. 1995; Wanders \& Peterson 1996),
the basic assumption in the dynamical method is considered to be robust.
In section 2, we discuss the method in detail.

All the previous estimates of nuclear mass have been made for type 1
Seyfert nuclei (hereafter S1s) and quasars because the dynamical method
needs both the flux and the velocity width of broad line emission.
This raises the question ^^ ^^ How massive are type 2 Seyfert
nuclei (hereafter S2s)  and are they similar to those of S1s ?''
Since the discovery of hidden BLR in the archetypical S2 nucleus of NGC 1068
by Antonucci \& Miller (1985), it has been considered that S2s are S1s 
in which the BLR as well as the central engine are hidden from
direct view (see, for a review Antonucci 1993). 
Taking this unified scheme into account, we may expect that there is no
systematic difference in the nuclear mass between S1s and S2s.
Miller \& Goodrich (1990; hereafter MG90) made a systematic study of hidden
BLRs of high-polarization S2s and found that the properties of the hidden BLRs
studied by polarized broad H$\alpha$ and  H$\beta$ emission 
are nearly  the same as those of S1s
in the following respects; equivalent widths, line widths, reddening, and 
luminosities (see also Tran 1995a). 
The intrinsic H$\beta$ luminosities\footnote{We adopt
a Hubble constant $H_0 = 50$ km s$^{-1}$ Mpc$^{-1}$ and a deceleration
parameter $q_0 = 0$ throughout this paper.}
of the S2s amount to $\sim 10^{43}$ erg s$^{-1}$.
MG90 adopted a typical H$\beta$ 
luminosity of $10^{43}$ erg s$^{-1}$ for S1s and thus reached the conclusion that
the intrinsic H$\beta$ luminosities are nearly the same between S1s and S2s.
However, the typical H$\beta$ luminosities  
of S1s are $\sim 10^{41}$ - $10^{42}$ erg s$^{-1}$  (Yee 1980; Blumenthal,
Keel, \& Miller 1982; Dahari \& De Robertis 1988). 
Therefore, the intrinsic H$\beta$ luminosities of S2s may be significantly more luminous
than those of S1s. However, although this suggests that there is a certain systematic
difference between S1s and S2s, it is noted that 
the intrinsic H$\beta$ luminosities of S2s depend on 
the estimates of the optical depth for electron
scattering, the degree  of {\it true} polarization, and the covering
factor of the scatterers (MG90). Thus the comparison of broad H$\beta$ luminosities 
between S1s and S2s must be made carefully.
After MG90, several new spectropolarimetric observations of S2s
have been published
(e.g., Tran, Miller, \&  Kay 1992;
Antonucci, Hurt, \& Miller 1994; Tran 1995a, 1995b).
New interpretations on the observed low polarizations have  been also presented
(Cid Fernandes \& Terlevich 1995; Heckman et al. 1995; Tran 1995c; 
Kishimoto 1996, 1997).
Therefore, it is interesting to revisit the comparison between
the hidden BLRs in S2s and the ordinary  BLRs in S1s and then
to estimate the dynamical masses  of S2 nuclei.

\section{DYNAMICAL MASS OF ACTIVE GALACTIC NUCLEI}

\subsection{The Dynamical Mass}

If we assume that the kinetic energy of the BLR is balanced
by the gravitational potential energy, then
the dynamical mass of a nucleus, $M_{\rm BH}$, is given by
\begin{equation}
M_{\rm BH} =	 \frac{\overline{v}^2 r} {2G} \simeq \frac{3 v^2 r} {2G}
\end{equation}
where  $\overline{v}$ is the mean velocity of the gas ($\overline{v}
= \sqrt{3} v = \sqrt{3}$FWHM where FWHM is the full width at half maximum),
$r$ is the size of
BLR, and $G$ is the gravitational constant [e.g., Padovani \& Rafanelli (1988;
hereafter PR88);
Padovani, Burg, \& Edelson (1990; PBE90), Koratkar \& Gaskell (1991b; KG91)].
Although various kinds of velocity widths are used in the literature,
they are related; 
FWZI/2 $\simeq$ HWZI  $\simeq \sqrt{3}$ FWHM where
FWZI and HWZI are full and half width at zero intensity, respectively
(Wandel \& Yahil 1985). It is however noted that either HWZI$_{\rm blue}$ or
HWZI$_{\rm red}$ is also sometimes used and they are usually different from
HWZI = FWZI/2 (e.g., PR88, PBE90). Since the maximum velocity is achieved 
at the innermost radius ($r_{\rm in}$) of BLR, the larger HWZI is usually 
adopted in the mass estimate.

Although the velocity width can be measured  directly from optical spectra,
it is difficult to estimate the innermost radius of BLR.
Therefore,  
this  is usually estimated
from photoionization considerations (e.g., PR88).
The ionization parameter, $U = Q({\rm H}) / (4 \pi r^2 n c)$,
gives an innermost radius,

\begin{equation}
r_{\rm in} = [Q({\rm H}) / 4 \pi c (U n)_{\rm in}]^{1/2}
\end{equation}
where $Q({\rm H})$ is the number of ionizing photons, $n$ is
the particle density, and $c$ is the light velocity in the vacuum.
From both observational and theoretical considerations,
PR88 adopted $(U n)_{\rm in} \simeq 10^{9.3}$ cm$^{-3}$ 
while PBE90 did $(U n)_{\rm in} \simeq 10^{9.5}$ cm$^{-3}$.
These give $r_{\rm in} \simeq 9.2\times10^{16} Q({\rm H})^{1/2}_{55}$ cm (PR88)
while $r_{\rm in} \simeq 3.2\times10^{16} Q({\rm H})^{1/2}_{55}$ cm (PBE90)
where $Q({\rm H})_{55} = Q({\rm H})/10^{55}$.
Since the variability estimated radii have been also taken into account in
PBE90, the estimate of $(U n)_{\rm in}$ by PBE90 is more reliable than
that by  PR88.
KG91 used for the first time
the distances of BLRs estimated  only by the cross correlation function (CCF) 
method for ten AGNs to estimate their nuclear masses (see also 
Gaskell \& Sparke 1986; Gaskell \& Peterson 1987).
If we compare the common objects in PR88, PBE90,
and KG91, we find that
the masses derived by KG91 tend to be more massive
than those by PR88 and PBE90.
The difference in the mass estimates of PR88, PBE90, and KG91 is due to
the two different lines and effective radii used in the analyses.
One would expect the mass estimate to be independent of the emission lines
used in the analyses if the gas was in Keplerian orbits.
Because of the proximity of the CIV line to the central massive engine,
this gas may not be in circular motion (e.g., Marziani et al. 1996), and
the widths of the CIV  may not be a true representation of the 
circular velocity at the CIV distance.
Given the uncertainties in the assumed velocity field and the assumed radii,
even though the mass estimates from PR88, PBE90, and KG91 seem to
differ by a factor of 3, the various methods are mostly consistent.

\subsection{The Dynamical Mass of Seyfert 2 Nuclei}

Hidden BLRs have been found in nine S2s (MG90; Tran 1995a,
1995b, 1995c). 
Using observational properties of the hidden BLRs,
we estimate nuclear masses  of the nine S2s. Here we use 
the following relation to estimate the dynamical mass,
\begin{equation}
M_{\rm BH}  \approx \frac{(\sqrt{3} {\rm  FWHM})^{2} r}{2G},
\end{equation}
Since the broad line emission of H$\beta$ was used in 
the previous estimates of $M_{\rm BH}$ for S1s (PR88; PBE90), 
we also use the {\it hidden} broad H$\beta$ emission for the
S2s in order to make a consistent
comparison of $M_{\rm BH}$ between S1s and S2s.
Since our main purpose is to compare the dynamical
masses of nuclei between S2s and S1s, the most important point
is to use {\it  a unique} dynamical method. 
Therefore we use equation (3) to estimate all the dynamical masses 
in this paper.
Following  PBE90, we use the inner radius of BLR by 

\begin{equation}
r  \simeq 3.6 \times 10^{16} Q({\rm H})_{54}^{1/2}
~ {\rm cm}
\end{equation}
where $Q({\rm H})$ is the number of hydrogen ionizing photons [$Q({\rm H})_{54}
= Q({\rm H})/10^{54}$].
Theoretically, there is a linear relationship between $Q({\rm H})$ and 
$L({\rm H\beta_b})$  where $L({\rm H\beta_{\rm b}})$ is the broad H$\beta$ luminosity
[e.g., $Q({\rm H}) \simeq 2.1 \times 10^{14} L({\rm H\beta_{\rm b}})$;
Peterson 1997]. However,  
the actual ultraviolet and X-ray observations of S1s give a relation,
$Q({\rm H}) \simeq 3.4 \times 10^{14} L({\rm H\beta_{\rm b}})^{0.95}$ (Padovani 1989;
PBE90). Adopting this relation, 
we estimate  the nuclear mass as

\begin{eqnarray}
M_{\rm BH} && \simeq 4.9 \times 10^7
Q({\rm H})_{54}^{1/2} [\sqrt{3} {\rm FWHM(H\beta_{\rm b})}_{6000}]^2 M_\odot \nonumber \\
&& \simeq 9.0 \times 10^6 L({\rm H\beta_{\rm b}})_{40}^{0.475}  
[\sqrt{3} {\rm FWHM(H\beta_{\rm b})}_{6000}]^2 M_\odot
\end{eqnarray}
where ${\rm FWHM(H\beta_{\rm b})}_{6000}$ is the FWHM of the broad H$\beta$
emission  in units of 6000 km s$^{-1}$ and $L({\rm H\beta_b})_{40}$
is the broad H$\beta$ luminosity in units of 10$^{40}$ erg s$^{-1}$.
Given both FWHM
and $L({\rm H\beta_{\rm b}})$, we are able to  estimate the dynamical mass of 
black holes in AGNs. 

In order to measure the FWHM of the hidden BLRs, MG90 made a correction 
for broadening by narrow-line components while Tran (1995a) gave the measured
values. There are only three S2s which are both in the MG90 and Tran (1995a)
samples. A comparison of these three S2s show that FWHM in two of these objects
are nearly similar. From these handful of objects we cannot accurately estimate 
the effect of the narrow line component on the measured FWHM, and hence we have 
adopted the FWHM determined by Tran (1995a) except for NGC 1068
for which we use the value determined by MG90.
we estimate that the uncertainty in the FWHM is no greater than 50\%,
which introduces a factor of 2.3 uncertainty in  the mass estimates.

In Table 1, we compare the FWHM for three S2s which were measured both
by MG90 and by Tran (1995a). Although the FWHM of NGC 7674 in MG90
is remarkably narrower than that in Tran (1995a), those of the remaining S2s
are nearly the same. Though it would be desirable to correct for the narrow-line
component, the FWHMs of S1s are usually not corrected for this effect.
Therefore we adopt the FWHM in Tran (1995a) except for NGC 1068 (MG90).
If the narrow line flux would contribute the polarized broad line flux,
we may overestimate the black hole mass. However, since it is difficult
to estimate this contribution accurately, we use the observed polarized
broad H$\beta$ fluxes given in Tran (1995a) in our analysis.

In order to estimate $M_{\rm BH}$ for S1s, 
we can use the observed FWHM(H$\beta_{\rm b}$)
and $L({\rm H\beta_{\rm b}})$ directly.
However, when we derive $M_{\rm BH}$ for S2s, 
the observed polarized, broad H$\beta$ luminosity, 
$L({\rm H\beta_{\rm b}})_{\rm p}$, must be converted to the intrinsic one
by using the following relation,
\begin{equation}
L({\rm H\beta_{\rm b}})_0 = L({\rm H\beta_{\rm b}})_{\rm p} \tau_{\rm es}^{-1} P^{-1} 
(\Delta\Omega/4 \pi)^{-1}
\end{equation}
where $\tau_{\rm es}$ is the optical depth for electron scattering\footnote{The 
numerical factor due to optical depth is $(1-e^{-\tau_{\rm es}})$
and thus equation (6) is valid for $\tau_{\rm es} \ll 1$. However, since
the factor $(1-e^{-\tau_{\rm es}}) \simeq 0.63$ for $\tau_{\rm es}$ = 1,
we use equation (5) approximately for $\tau_{\rm es} \lesssim 1$.}, $P$ is the 
polarization, and $\Delta\Omega$ is the covering factor of the scatterers in 
steradian (MG90).
It is observed in most cases that the degree of polarization of
broad wings of Balmer emission lines is larger than that of nonstellar 
continuum (e.g., Tran 1995a). This suggests a substantial contribution from a
{\it unpolarized} continuum, which should be corrected for in the estimate
of intrinsic polarization (Tran 1995c). Therefore, the degree of polarization
used in our analysis is from Table 1 of Tran (1995c) where these corrections
have been accounted. 
The covering factor, $\Delta\Omega / 4 \pi$, can be estimated using the full opening 
angle of the narrow line regions, $\theta_{\rm open}$:
$\Delta\Omega / 4 \pi = 1 -$ cos$(\theta_{\rm open}/2)$ (MG90).

The average value of $\theta_{\rm open}$ of 12 
Seyferts studied by Schmitt \& Kinney (1996) is 58$^\circ \pm 18^\circ$
(see also Pogge 1989; Wilson \& Tsvetanov 1994),
corresponding to $\Delta\Omega / 4 \pi \simeq$ 0.125 on the average.
Although there seems a factor of 2 scatter in the opening angle, 
this introduces  only a factor of 1.4 ($\simeq 2^{0.475}$) uncertainty 
in the mass estimates. 
Hence, we adopt  
$\Delta\Omega / 4 \pi$ = 0.1 for all the objects.
Since $\tau_{\rm es}$ is usually difficult to estimate, 
as  it depends on electron density, density distribution,
etc, therefore,  we regard $\tau_{\rm es}$  as a free parameter.
The effect of $\tau_{\rm es}$ will be discussed later.

Finally we summarize the total uncertainty in the dynamical mass estimates.
The largest uncertainty comes from the estimate of FWHM. As mentioned before,
the uncertainty in the FWHM introduces a factor of 2.3 uncertainty in the
mass estimate. Since this uncertainty is due to the correction for the narrow line
component, this results in overestimates of the mass.
Other uncertainty comes from both the estimates of
$Q$(H$^0$) and  $\Delta\Omega/4\pi$.
Though the uncertainty in $Q$(H$^0$) can be as large as a factor of 3
because of the various assumptions used in calculating this parameter,
this uncertainty introduces a factor of 1.7 uncertainty in the mass estimate.
The uncertainty in $\Delta\Omega/4\pi$ also introduces a factor of 1.4 uncertainty.
Thus there is a factor of 5.5 uncertainty
in total in the dynamical mass estimate.

\subsection{Results}

In Table 2, we present our results.
We obtain an  average nuclear mass, log$(M_{\rm BH} / M_\odot) 
\simeq  8.00 \pm 0.51 - 0.475 {\rm log}(\tau_{\rm es}/1)$.
If $\tau_{\rm es} = 0.1$, log$(M_{\rm BH} / M_\odot)
\simeq  8.47 \pm 0.51$.
We compare our results with those for S1s studied by PR88 and PBE90.
In both PR88 and PBE90, they used either HWZI$_{\rm blue}$ or
HWZI$_{\rm red}$. In order to make consistent  comparisons of our
results with theirs, we evaluate nuclear  masses of  their S1s
using the same method as that of our analysis [equation (5)].
Since the FWZI is given for 30 S1s in PR88 and 25 S1s in PBE90,
we can estimate nuclear masses only for these objects, where in  equation (5)
we use FWZI/2 instead of $\sqrt{3}$FWHM.
The average masses are compared in Table 3.
The frequency distributions of log$(M_{\rm BH} / M_\odot)$ 
are compared in Fig. 1.
We also compare the $M_{\rm BH}$ - $L({\rm H\beta_{\rm b}})$ relation
between S1s and S2s
for the two cases; $\tau_{\rm es} = 0.1$ and $\tau_{\rm es} = 0.1$
in Fig. 2.
The relation for S2s follows nearly the same relation for S1s
for both cases. However, if $\tau_{\rm es} = 0.1$, the S2s tend to be
more luminous and more massive than the S1s.

We examine the similarity of 
the frequency distribution of masses between S2s and S1s shown in Fig. 1.
We now adopt the null hypothesis that the S1 and S2 masses
come from the same underlying distribution.
Applying  Kormogrov-Smirnov statistical test, 
we obtain the probability of randomly selecting the masses from the same underlying
distribution is; (1) 27.3\% for S2s($\tau_{\rm es}=1$) vs. S1s(PR88),
(2) 98.2\% for S2s($\tau_{\rm es}=1$) vs. S1s(PBE90),
(3) 0.26\% for S2s($\tau_{\rm es}=0.1$) vs. S1s(PR88), and
(4) 10.7\% for S2s($\tau_{\rm es}=0.1$) vs. S1s(PBE90).
All but the third case show that the dynamical masses of S2s are 
similar to those of S1s.
Given the uncertainty in the dynamical mass estimate (a factor of 5.5),
we cannot conclude  that the  S2s have more massive nuclei than the S1s
even for the third case.
However, we should keep in mind that 
the S2 masses tend to be more massive with decreasing $\tau$.
Since low optical depths have been often adopted in previous works,
it will become important to estimate $\tau_{\rm es}$ accurately for many S2s.

Since our S2 sample is taken from the literature which presents 
the spectropolarimetry,
the S2s studied here maybe biased to show higher polarizations. However, 
a comparison of the apparent and absolute magnitudes of S2s
with (7 objects) and without (42 objects) hidden BLRs show that
there is no significant difference between the two samples.
Hence it is unlikely that the S2s with polarized H$\beta$ are 
intrinsically more luminous and therefore more massive.
This seems reasonable because the polarization
depends solely on the physical conditions of electron scattering regions
as well as on the viewing angle toward them.
Further, the three samples compared here have the similar average redshifts;
$z = 0.028 \pm 0.019$ for the S2s, $0.036 \pm 0.022$ for the S1s (PR88),
and $0.024 \pm 0.017$ for the S1s (PBE90). 
Therefore, it  is not expected that there are strong observational  biases
in the three samples although our samples are not {\it complete} ones 
in any sense except for the CfA sample studied 
by PBE90 (see Huchra \& Burg 1992).

\section{DISCUSSION}

We have derived the dynamical masses of S2 nuclei. Comparing them with
those of S1s, we have obtained that the nuclear masses of S2s are similar
to those of S1s provided that $\tau_{\rm es} \sim 1$ while 
more massive if $\tau_{\rm es} \ll 1$.
For example, if $\tau_{\rm es} \sim 0.1$, 
the nuclear masses of S2s would be systematically 
larger by an order of magnitude than those of S1s. 
We now consider physical conditions in the electron scattering regions, and
we discuss some implications for the unified model of Seyfert nuclei.

Low optical depths have been adopted
in previous works based on spectropolarimetry of S2s;
e.g., $\tau_{\rm es} \simeq$ 0.05 - 0.1 (Antonucci \& Miller 1985;
MG90; Miller et al. 1991). 
As discussed by MG90, if $\tau_{\rm es} \ll  1$, we would observe many AGN with
polarizations higher than 50\% 
provided that  the half opening angle of the ionization cone
$\theta_{\rm c} = \theta_{\rm open}/2
 \sim 30^\circ$ as is observed for many S2s (Pogge 1989;
Wilson \& Tsvetanov 1994;
Schmitt \& Kinney 1996). However, the highest polarization observed so far
is 16\% (NGC 1068: MG90; Antonucci et al. 1994)
and the typical polarization is only several percent 
for the other S2s (MG90; Tran 1995a).
The observed polarization is lower than expected,
even after  the effect of dilution due to the unpolarized continuum radiation
is taken into account (Cid Fernandes \& Terlevich 1995; Heckman et al 1995;
Tran 1995c; Tran, Cohen \& Goodrich 1995; see also  Kishimoto 1996).
Therefore, the optical thick condition cannot be ruled out entirely
at present. 

We now discuss the physical characteristics of the electron scattering region.
The optical depth for electron scattering is estimated by 

\begin{equation}
\tau_{\rm es} \sim \sigma_{\rm T} {\overline{n}_{\rm e}} l_{\rm eff} 
\sim \sigma_{\rm T} N_{\rm e}
\end{equation}
where $\sigma_{\rm T}$ is the Thomson cross section, 0.66$\times 10^{-24}$ cm$^2$,
${\overline{n}_{\rm e}}$ is the average electron density in the scattering region,
$l_{\rm eff}$ is the effective path length, and $N_{\rm e}$ is the electron 
column density.
We obtain $N_{\rm e} \sim 10^{24}$ cm$^{-2}$ for $\tau_{\rm es} \sim 1$ while
$N_{\rm e} \sim 10^{23}$ cm$^{-2}$ for $\tau_{\rm es} \sim 0.1$.
If the kinetic temperature of free electrons is as high as $\sim 10^6$ K,
significant line broadening would  occur due to the scattering 
(Antonucci \& Miller 1985; MG90).
Since $v \sim 2000 (T_{\rm e}/10^5 {\rm K})$
km s$^{-1}$, it seems reasonable to assume  $T_{\rm e} \sim 10^5$ K at most
(see also Miller et al. 1991).

If the gas in the electron scattering region is in pressure equilibrium with 
the BLR gas ($n_{\rm e} \sim 10^9$ cm$^{-3}$ and $T_{\rm e} \sim 10^4$ K; 
Osterbrock 1989), then the average electron density in the scattering
region is ${\overline{n}_{\rm e}} \sim 10^8$ cm$^{-3}$,  and 
the effective path lengths are $\sim 10^{16}$ cm and $\sim 10^{15}$ cm
for $\tau_{\rm es} \sim 1$  and $\sim 0.1$, respectively.
Since we observe the BLR through the scatterers, 
the scattering regions are located outside
the BLRs and thus the radial distance of scatterers is
$r_{\rm e} > 10^{16}$ cm.
In the case of NGC 1068, it is observed that the scattering regions 
are extended to $\sim$ 100 pc from the nucleus (Capetti et al. 1995a, 1995b).
This large value was indeed suspected from the photoionization consideration by
Miller et al. (1991). 
It is therefore considered that the electron scattering regions are located
at $r_{\rm e} \sim 10^{16}$ - $10^{20}$ cm. 
The effective radius of scatterers may be different from object to object
because it depends also on how we observe the dusty tori in AGNs (i.e., viewing angle
dependent; cf. MG90, Miller et al. 1991; Kishimoto 1996, 1997). 

It is worth noting that the above physical conditions are quite similar to those of 
warm absorbers probed by X-ray spectroscopy of type 1 AGNs (Halpern 1984; Netzer 1993;
Nandra \& Pounds 1992, 1994;
Ptak et al. 1994; Reynolds \& Fabian 1995; Reynolds et al. 1995; Otani et al. 1996;
Reynolds 1997).
However, the electron column density inferred in this study, 
$N_{\rm e} \sim 10^{23}$ - $10^{24}$ cm$^{-2}$,
is higher by one to two orders of magnitude than those of warm absorbers,
$\sim 10^{22}$ cm$^{-2}$.
This seems to be  inconsistent with the strict unified model of Seyfert
nuclei (Antonucci \& Miller 1985).
We may, however, consider the higher column densities in the S2s 
as  due to the  effect of multiple scattering if 
$\tau_{\rm es} \sim 1$.
Another interpretation may be that
S2s are gas-richer systematically  than S1s (cf. Heckman et al. 1989; Taniguchi 1997).

\vspace{0.5cm}

We  would like to thank Makoto Kishimoto, Youichi Ohyama,
and Takashi Murayama for useful discussion.
We also thank the anonymous referee for his/her many useful comments and
suggestions which improved this paper significantly.
This work was financially supported in part by Grant-in-Aids for the Scientific
Research (No. 07044054) of the Japanese Ministry of
Education, Culture, Sports, and Science.

%------------------------------------------------------------------------------
\newpage

\begin{table}
\caption{Comparison of the FWHM(H$\beta$)\tablenotemark{a} ~ between MG90 and Tran(1995a)}
\vspace{5mm}
\begin{tabular}{ccc}
\tableline
\tableline
Galaxy  & MG90 & Tran (1995a)  \\
\tableline
Mrk 3 & 5400 & 6000$\pm$500 \\
Mrk 463E & 3000 & 2770$\pm$180 \\
NGC 7674 & 1500 & 2830$\pm$150 \\
\tableline
\tablenotetext{a}{In units of km s$^{-1}$.}
\end{tabular}
\end{table}

%------------------------------------------------------------------------------

\begin{table}
\caption{The dynamical masses of Seyfert 2 nuclei}
\vspace{5mm}
\begin{tabular}{ccccccc}
\tableline
\tableline
Object & FWHM(H$\beta_{\rm b}$) & $L({\rm H}\beta_{\rm b})_{\rm p}$\tablenotemark{a} 
& $P$\tablenotemark{b}
& $M_{\rm BH} (\tau_{\rm es}=1)$ & $M_{\rm BH} (\tau_{\rm es}=0.1)$ & Ref.\tablenotemark{c} \\
 & (km s$^{-1}$) & (erg s$^{-1}$) & (\%) & ($M_{\odot}$) & ($M_{\odot}$) &  \\
\tableline
NGC 1068 & 3030 & $1.11\times10^{39}$ & 16 & $1.73\times10^7$ & $5.16\times10^7$ & 1, 2 \\
NGC 7212 & 5420 & $3.40\times10^{39}$ & 22 & $8.09\times10^7$ & $2.42\times10^8$ & 2  \\
NGC 7674 & 2830 & $5.36\times10^{39}$ & 8 &  $4.43\times10^7$ & $1.32\times10^8$ & 2  \\
Mrk 3    & 6000 & $1.00\times10^{40}$ & 20 & $1.73\times10^8$ & $5.17\times10^8$ & 2  \\
Mrk 348  & 9350 & $2.95\times10^{39}$ & 35\tablenotemark{d} & $1.81\times10^8$ & $5.39\times10^8$ & 2 \\
Mrk 463E & 2770 & $3.99\times10^{40}$ & 10 & $9.89\times10^7$ & $2.95\times10^8$ & 2 \\
Mrk 477  & 4130 & $7.85\times10^{40}$ & 2 &  $6.66\times10^8$ & $1.96\times10^9$ & 2 \\
Mrk 1210 & 3080 & $2.22\times10^{39}$ & 15 & $2.56\times10^7$ & $7.63\times10^8$ & 2 \\
Was 49   & 5860 & $3.30\times10^{40}$ & 20 & $2.91\times10^8$ & $8.69\times10^8$ & 2 \\
\tableline
\tablenotetext{a}{Luminosity of polarized, broad H$\beta$ emission.}
\tablenotetext{b}{Intrinsic polarization corrected for the unpolarized continuum
emission as well as interstellar polarization taken from Tran (1995c).}
\tablenotetext{c}{1. Miller \& Goodrich 1990; 2. Tran 1995a, 1995c.}
\tablenotetext{d}{The intrinsic polarization for H$\alpha$ emission.}
\end{tabular}
\end{table}

%------------------------------------------------------------------------------

\begin{table}
\caption{Comparison of the average dynamical masses between S2s and S1s}
\vspace{5mm}
\begin{tabular}{ccc}
\tableline
\tableline
Sample & Number & $M_{\rm BH}$ \\
 &  & ($M_{\odot}$) \\
\tableline
S2 ($\tau_{\rm es}=1$) & 9 & 8.00$\pm$0.51 \\
S2 ($\tau_{\rm es}=0.1$) & 9 & 8.47$\pm$0.51 \\
S1 (PR88) & 30 & 7.89$\pm$0.57 \\
S1 (PBE90) & 25 & 7.48$\pm$0.63 \\
\tableline
\end{tabular}
\end{table}

%------------------------------------------------------------------------------

\newpage

\figcaption{Frequency distributions of  black hole masses ($M_{\rm BH} / \rm M_{\odot}$)
for the S2s studied here, S1s studied by both PR88 and PBE90. As for the S2s,
we show the results for $\tau_{\rm es}$ = 1 and $\tau_{\rm es}$ = 0.1 separately.
\label{fig1}}

\figcaption{The $M_{\rm BH}$ - $L_{\rm H\beta_{\rm b}}$ relation is 
compared between S1s and S2. The left panel shows the case of $\tau_{\rm es} = 1$
for S2s while the right panel shows the case of $\tau_{\rm es} = 0.1$.
The strait lines show the regression line for the S1 data (PR88 and PBE90).
\label{fig2}}


\newpage

\begin{references}
\reference{1}{Antonucci, R. 1993, ARA\&A, 31, 473}
\reference{1}{Antonucci, R., Hurt, T., \& Miller, J. 1994, ApJ, 430, 210}
\reference{1}{Antonucci, R., \& Miller, J. S. 1985, ApJ, 297, 621}
\reference{1}{Blandford, R. D. 1990, in Active Galactic Nuclei, ed. R. D. Blandford, 
              H. Netzer, L. Woltier (Splinger-Verlag), 161}
\reference{1}{Blandford, R. D., \& McKee, C. F. 1982, ApJ, 255, 419}
\reference{1}{Blumenthal, G. R., Keel, W. C., \& Miller, J. S. 1982, ApJ, 257, 499}
\reference{1}{Capetti, A., Axon, D. J., Macchetto, F., Sparks, W. B.,
              \& Boksenberg, A. 1995a, ApJ, 446, 155}
\reference{1}{Capetti, A., Macchetto, F., Axon, D. J., Sparks, W. B.,
              \& Boksenberg, A. 1995b, ApJ, 452, L87}
\reference{1}{Cid Fernandes, R. Jr., \& Terlevich, R. 1995, MNRAS, 272, 423}
\reference{1}{Clavel, J., et al. 1991, ApJ, 336, 64}
\reference{1}{Dahari, O. V., \& De Robertis, M. M. 1988, ApJS, 67, 249}
\reference{1}{de Vaucouleurs, G., de Vaucouleurs, A., Corwin, H. G., Jr., Buta, R. J.,
              Paturel, G., \& Forqu\'e, P. 1991, Third Reference Catalogue of Bright
              Galaxies (Springer-Verlag)}
\reference{1}{Dibai, E. A. 1981, Soviet Astr., 24, 389}
\reference{1}{Dibai, E. A. 1984, Soviet Astr., 28, 245} 
\reference{1}{Gaskell, C. M. 1988, ApJ, 325, 114}
\reference{1}{Gaskell, C. M., \& Peterson, B. M. 1987, ApJS, 65, 1}
\reference{1}{Gaskell, C. M., \& Sparke, L. S. 1986, ApJ, 305, 175}
\reference{1}{Haehnelt, M. G., \& Rees, M. J. 1993, MNRAS, 263, 168}
\reference{1}{Halpern, J. P. 1984, ApJ, 281, 90}
\reference{1}{Heckman, T. M., Beckwith, S., Blitz, L., Skrutskie, M., 
             \& Wilson, A. S. 1989, ApJ, 342, 735}
\reference{1}{Heckman, T. M., et al. 1995, ApJ, 453, 549}
\reference{1}{Heisler, C. A., \& Vader, J. P. 1994, AJ, 107, 35}
\reference{1}{Huchra, J., \& Burg, R. 1992, ApJ, 393, 90}
\reference{1}{Joly, M., Collin-Souffrin, S., Masnou, J. L., \& Nottale, L. 1985, A\&A, 152, 282}
\reference{1}{Kay, L. E. 1994, APJ, 430, 196}
\reference{1}{Kishimoto, M. 1996, ApJ, 468, 606}
\reference{1}{Kishimoto, M. 1997, ApSS, 248, 277}
\reference{1}{Koratkar, A. P. \& Gaskell, C. M. 1991a, ApJS, 75, 719}
\reference{1}{Koratkar, A. P. \& Gaskell, C. M. 1991b, ApJ, 370, L61 (KG91)}
\reference{1}{Koratkar, A. P. \& Gaskell, C. M. 1991c, ApJ, 375, 85}
\reference{1}{Korista, K. T., et al. 1995, ApJS, 97, 285}
\reference{1}{Marziani, P., Sulentic, J. W., Dultzin-Hacyan, D., Calvani, M., \&
              Moles, M. 1996, ApJS, 104, 37}
\reference{1}{Miller, J. S., \& Goodrich, R. W. 1990, ApJ, 355, 456 (MG90)}
\reference{1}{Miller, J. S., Goodrich, R. W., \& Mathews, W. G. 1991, ApJ, 378, 47}
\reference{1}{Nandra, K., \& Pounds, K. 1994, Nature, 359, 215}
\reference{1}{Nandra, K., \& Pounds, K. 1994, \mnras, 268, 405}
\reference{1}{Nelson, C. H., \& Whittle, M. 1995, ApJS, 99, 67}
\reference{1}{Netzer, H. 1993, ApJ, 411, 594}
\reference{1}{Osterbrock, D. E. 1989, Astrophysics of Gaseous Nebulae and 
              Active Galactic Nuclei (University Science Book)}
\reference{1}{Otani, C., et al. 1996, \pasj, 48, 211}
\reference{1}{Padovani, P. 1989, 209, 27}
\reference{1}{Padovani, P., Burg, R., \& Edelson, R. A. 1990, ApJ, 353, 438 (PBE90)} 
\reference{1}{Padovani, P., \& Rafanelli, P. 1988, A\&A, 205, 53 (PR88)}
\reference{1}{Peterson, B. M. 1993, PASP, 105, 247}
\reference{1}{Peterson, B. M. 1997, An Introduction to Active Galactic Nuclei
              (Cambridge University Press), 117}
\reference{1}{Pier, E., \& Krolik, J. 1993, ApJ, 418, 673}
\reference{1}{Pogge, R. W., 1989, ApJ, 345, 730}
\reference{1}{Ptak, A., Yaqoob, T., Serlemitsos, P., \& Mushotzky, R. 1994, \apj, 436, L31}
\reference{1}{Rees, M. J. 1984, ARA \& A, 22, 471}
\reference{1}{Reshetnikov, V. P. 1987, Astrofizika, 27, 283}
\reference{1}{Reynolds, C. S. 1997, MNRAS, 286, 513}
\reference{1}{Reynolds, C. S., \& Fabian, A. C. 1995, MNRAS,  273, 1167}
\reference{1}{Reynolds, C. S., Fabian, A. C., Nandra, K., Inoue, H., Kunieda, H., \&
              Iwasawa, K. 1995, MNRAS, 277, 901}
\reference{1}{Robinson, A. 1994, in ASP Conf. Ser. 69, Reverberation Mapping of the Broad-Line
              Region of Active Galactic Nuclei, ed. P. M. Gondhalekar, K. Horne, \&
              B. M. Peterson (San Francisco: ASP), 147}
\reference{1}{Salzer, J. J., Moody, J. W., Rosenberg, J. L., 
Gregory, S. A., \& Newberry, M. V. 1995, AJ, 109, 2376}
\reference{1}{Sandage, A., \& Tammann, G. A. 1981, A Revised 
Shapley-Ames Catalog of Bright Galaxies (CARNEGIE INSTITUTION OF WASHINGTON)}
\reference{1}{Schmitt, H. R., \& Kinney, A. L. 1996, ApJ, 463, 498}
\reference{1}{Setti, G., \& Woltier, L. 1966, ApJ, 144, 838}
\reference{1}{Taniguchi, Y. 1997, ApJ, 487, L17}
\reference{1}{Tran, H. D. 1995a, ApJ, 440, 565}
\reference{1}{Tran, H. D. 1995b, ApJ, 440, 578}
\reference{1}{Tran, H. D. 1995c, ApJ, 440, 597}
\reference{1}{Tran, H. D., Cohen, M. H., \& Goodrich, R. W. 1995, AJ, 110, 2597}
\reference{1}{Tran, H. D., Miller, J. S., \& Kay, L. E. 1992, ApJ, 397, 452}
\reference{1}{Wandel, A., \& Mushotzky, R. F. 1986, ApJ, 306, L61} 
\reference{1}{Wandel, A., \& Yahil, A. 1985, ApJ, 295, L1}
\reference{1}{Wanders, I., et al. 1995, ApJ, 453, L87}
\reference{1}{Wanders, I., \& Peterson, B. M. 1996, ApJ, 466, 174}
\reference{1}{Wanders, I., \& Peterson, B. M. 1997, ApJ, 477, 990}
\reference{1}{Wilson, A. S., \& Tsvetanov, Z. I. 1994, AJ, 107, 1227}
\reference{1}{Woltier, L. 1959, ApJ, 130, 38}
\reference{1}{Yee, H. K. C. 1980, ApJ, 241, 894}
\end{references}
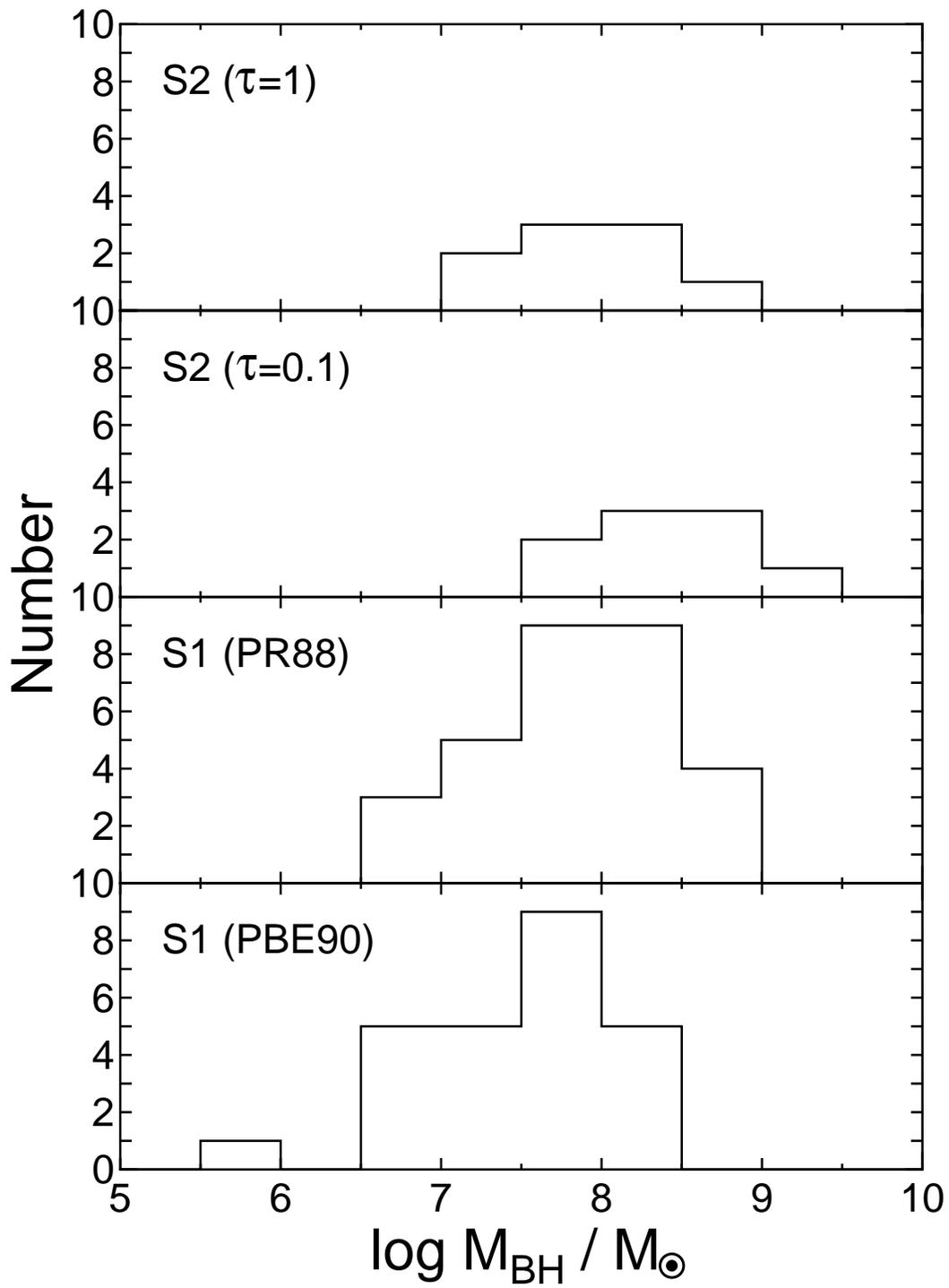
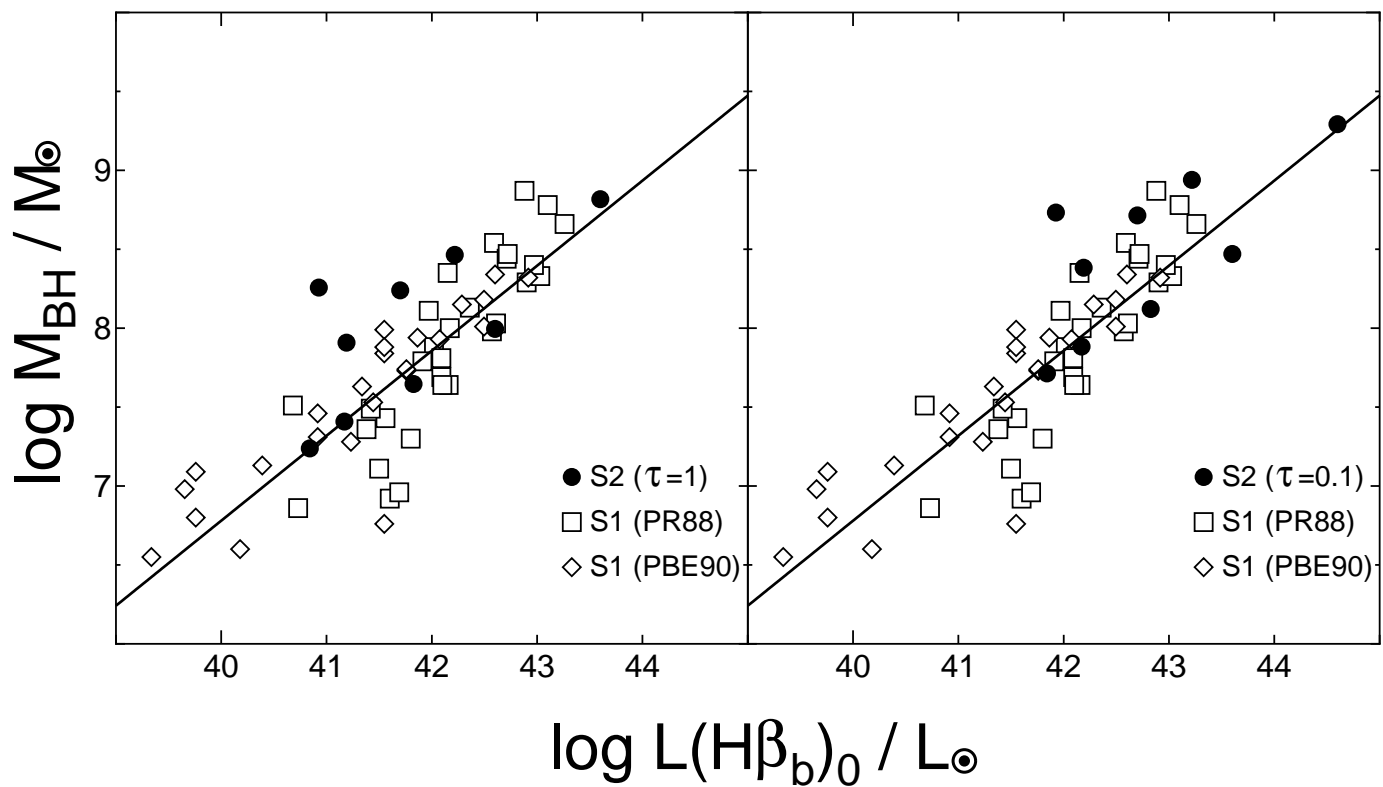
\end{document}